\documentclass[proceedings]{JHEP}
\usepackage{epsfig}                   % please use epsfig for figures
\renewcommand{\nc}{\newcommand}
\long\def\omit#1{}

\nc{\be}{\begin{equation}}
\nc{\ee}{\end{equation}}
\nc{\beqa}{\begin{eqnarray}}
\nc{\eeqa}{\end{eqnarray}}
\nc{\bea}{\begin{eqnarray}}
\nc{\eea}{\end{eqnarray}}

\nc{\Fo}{{{}^{(1)}\!F}}
\nc{\Fos}{{{}^{(2)}\!F}}
\nc{\Fot}{{{}^{(3)}\!F}}
\nc{\Fota}{{{}^{(4)}\!F}}
\nc{\bra}{\langle}\nc{\ket}{\rangle}

\nc{\za}{\alpha} \nc{\zb}{\beta} \nc{\zg}{\gamma} \nc{\zd}{\delta}
\nc{\ze}{\varepsilon} \nc{\zf}{\phi} \nc{\zv}{\varphi} \nc{\zh}{\chi}
\nc{\zk}{\kappa} \nc{\zl}{\lambda} \nc{\zm}{\mu} \nc{\zn}{\nu}
\nc{\zo}{\omega} \nc{\zp}{\pi} \nc{\zr}{\rho} \nc{\zs}{\sigma} 
\nc{\zt}{\tau} \nc{\zz}{\zeta}\nc{\zi}{\eta}
\nc{\slh}{\widehat{sl}}

\nc{\bl}{\bar{\lambda}}
\nc{\bL}{\bar{\Lambda}}
\nc{\zG}{\Gamma}  \nc{\zD}{\Delta} \nc{\zF}{\Phi}
\nc{\zL}{\Lambda} \nc{\zO}{\Omega} \nc{\zS}{\Sigma}

\font\tenmsb=msbm10 scaled \magstep1
\font\sevenmsb=msbm7 scaled \magstep1
\font\fivemsb=msbm5 scaled \magstep1\newfam\msbfam
\textfont\msbfam=\tenmsb
\scriptfont\msbfam=\sevenmsb
\scriptscriptfont\msbfam=\fivemsb
\def\Bbb#1{{\fam\msbfam\relax#1}}

\nc{\calA}{{\cal A}}
\nc{\calH}{{\cal H}}
\nc{\calB}{{\cal B}}
\nc{\calC}{{\cal C}}
\nc{\calE}{{\cal E}}
\nc{\calI}{{\cal I}}
\nc{\calM}{{\cal M}}
\nc{\calN}{{\cal N}}
\nc{\calS}{{\cal S}}
\nc{\calV}{{\cal V}}

\nc{\jb}{
\bar{j}}\nc{\ib}{\bar{i}}\nc{\kb}{\bar k}
\nc{\llangle}{\langle\!\langle}
\nc{\rrangle}{\rangle\!\rangle}
\nc{\oh}{{1\over 2}}
\nc{\tr}{{\rm tr\,}}\nc{\mod}{{\rm mod\,}}
\nc{\tV}{\tilde{V}}

\input amssym.def
\input amssym.tex

\def\youngone{{ \hbox{\epsfxsize=1.5mm\epsfbox{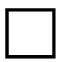}} }}
\def\youngtwo{ \hbox{\raise -0.5mm\hbox{\epsfxsize=1.5mm\epsfbox{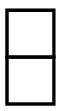}} }}

\conference{Nonperturbative Quantum Effects 2000}

\title{ BCFT: from the boundary to the bulk}

\author{V.B. Petkova 
$\!\!$\thanks{
permanent address:
Institute for Nuclear Research and Nuclear Energy,
72 Tzarigradsko Chaussee,
1784 Sofia, Bulgaria;
petkova@inrne.bas.bg}
\ and J.-B. Zuber\\
School of Computing and Mathematics\\
University of Northumbria\\
NE1 8ST Newcastle upon Tyne, UK\\
E-mail: \email{valentina.petkova@unn.ac.uk}\\
and\\
SPhT, CEA Saclay\\
91191 Gif-sur-Yvette, France\\
E-mail:  \email{zuber@spht.saclay.cea.fr}}

\abstract{
The study of boundary conditions in rational conformal field
theories is not only physically important. It also reveals a lot
on the structure of the theory ``in the bulk''. The same graphs
classify both the torus and the cylinder partition functions and
provide data on their hidden ``quantum symmetry''.  The Ocneanu
triangular cells -- the $3j$-symbols of these symmetries, admit
various interpretations and make a link between different
problems.

}

\begin{document}
\section{Introduction}

Ten years after the work of Cardy \cite{Ca89}, Boundary Conformal 
Field Theory (BCFT) is experiencing a  renewal of interest. This 
is motivated by applications to string and brane theory, and to 
problems of statistical mechanics and condensed matter. In these 
situations, it may be important to have an a priori knowledge of
the possible boundary conditions compatible with conformal 
invariance, and to master the algebra of boundary fields etc. 
There is, however, another reason to be interested in BCFT: 
as we want to explain in this note, there is much to learn 
on the general structure and the quantum symmetries of a CFT from 
the study of its properties in the presence of a boundary.
After a brief review of notations and general aspects of
rational CFT, we shall discuss how boundary conditions may be
systematically classified in terms of 
non-negative integer valued matrix representations 
of the fusion algebra, or equivalently in terms of graphs generalising 
the ADE Dynkin diagrams (see \cite{BPPZ} for more details). We shall see 
 how the ``cells'' -- a concept introduced by
Ocneanu and  associated with these graphs --
 determine many properties
of the BCFT or of the associated lattice models 
 and, in particular,  are at the heart of the quantum symmetry of
 the 
 CFT described by an algebraic structure called weak  $C^*$-Hopf
 algebra.  A more detailed presentation will appear in \cite{PZ4}.

%%%%%%%%%%%%%%%%%%%%%%%%%%%%%%%%%%%%%%%%%%%%%%%%%%%%%%%%%%%%%%%%%%%%%%%%%%%%%%

\section{General set-up}
 A rational conformal field theory (RCFT) is generally described
by data of different nature:\\
%%%%%%%%%%
\noindent
$\bullet$  {\bf Chiral data:} 
%--------------------------- 
Chiral data  specify the  chiral
algebra ${\goth A}$, e.g., the   Virasoro algebra itself, or a
${\cal  W}$ algebra, a current algebra $\widehat{ g}\,$ etc, and
its finite set $\calI$ of irreducible representations,
$\calV_i\,, \ i\in \cal  I\,$; 
notations are such that $i=1$ labels
the identity (vacuum)  representation and $i^*$ the conjugate of
${\cal V}_i$.
  The  fusion rule multiplicities  $N_{ij}{}^k$, $
\calV_i \star \calV_j =\oplus_k\, N_{ij}{}^k\,  \calV_k\,, $ are
assumed to be given by the Verlinde formula,
\be 
N_{ij}{}^k=\sum_{\ell\in{ I}}{S_{il}\, S_{jl}
\left(S_{kl}\right)\!{}^*
\over S_{1l}} %\;\in\; {\Bbb Z_{\ge 0}}
\, ,
%{\Bbb N}\,,
\label{Verl} 
\ee 
with $S$ the symmetric, unitary matrix of the modular
transformations of the characters $\chi_i(\tau)=\tr
e^{2i\pi\tau(L_0-c/24)}$, $\chi_i(\tau)=\sum_{j\in \calI} S_{ij}
\chi_j(-1/\tau)$.
The nonnegative integers $N_{ij}{}^k$ give the dimensions 
of linear spaces of chiral vertex operators (CVO)
\be 
\phi_{ij;t,I}^k(z):\ \calV_i \otimes \calV_j \to \calV_k \,, \ \
%t=1,2,\dots N_{ij}{}^k,\ 
z\in {\Bbb C}\,,
\ee 
with a finite basis label  $t=1,2,\dots N_{ij}{}^k\,,$
%with $t=1,2,\dots N_{ij}{}^k\,$ a basis label
and  $I$ labelling descendents in $\calV_i$.
The chiral data finally include the knowledge of the duality
matrices: the genus $0$ fusing $F$ and braiding $B(\pm)$
matrices, and the matrix $S(j)$, which gives the modular
transformation of $1$-point conformal block $\bra \phi_{ji}^i
\ket$ on the  torus. These matrices satisfy a set of
consistency relations: pentagon, hexagon and  torus identities
\cite{MS1}. 
%In particular, the matrix $F$ satisfies a pentagon
%identity expressing the associativity of the fusion of CVO.

A typical example is provided by the chiral algebra ${\goth
A}=\slh(2)_k$, the affine algebra at level $k$, for which
$\calI=\{1,2,\cdots,k+1\}$, $S_{ij}=\sqrt{2/(k+2)}$ $\sin{\pi
ij/(k+2)}$,  and the $F$ are, up to a gauge transformation, the
quantum $6j$ symbols \cite{KiRe}.

Those are the basic ingredients of algebraic nature in the
construction of a RCFT. In the following, they are supposed to be
known.

%%%%%%%%%% 
%\medskip
\noindent
$\bullet$  {\bf Spectral and OPE data:}
%--------------------------------------
The physical spectrum of a RCFT {\it in the bulk}, i.e. on a
closed Riemann surface, is described  by irreducible
representations of {\it two} copies of  the chiral algebra. Thus
the Hilbert space is 
decomposed according to
\be 
{\cal H}_P=\oplus_{_{j,\jb\,\in{ \calI}}}\,
 Z_{j\, \bar{j}}\, \calV_j\otimes \bar{\calV}_{\bar j}\,.
 \label{Spec}
\ee 
The integer multiplicities $Z_{j\,\bar{j}}$ (with $Z_{11}=1$,
expressing the unicity of the vacuum) are conveniently encoded in
the modular invariant torus partition function
\be 
Z(\tau)=
\sum_{j,\jb\,\in{ \calI}}\,  Z_{j\jb}\ \chi_j({\tau})\,
\chi_{\jb}(\tau)^* \,.
\label{partfn}  
\ee 
 The particular case where $Z_{ij}=\delta_{ij}$ will be referred
to as the diagonal theory, the others as ``non-diagonal''.

On the other hand, 
the  physical fields and their correlators factorise
\omit{\be 
\langle \cdots \Phi_{(i,\ib)}(z,\bar{z}) \cdots \rangle
 =
 \dots \sum_{j, \jb, k, \bar{k}, t,\bar{t}}\,  
d_{(i,\ib) (j,\jb)}^{(k,\bar{k});t, \bar{t}}\cdots
\langle \cdots \phi_{i,j;t}^k(z)\,  \cdots \rangle\ \otimes
\langle \cdots \phi_{\ib,\jb;\bar{t}}^{\bar{k}}(\bar{z})\, \cdots \rangle
\ee 
}
\be
\Phi_{(i,\ib)}(z,\bar{z})=\!\!\! \sum_{j, \jb, k, \bar{k}, t,\bar{t}}\,
\!\!\! d_{(i,\ib) (j,\jb)}^{(k,\bar{k});t, \bar{t}} \,
\phi_{i,j;t}^k(z)\, \otimes
\phi_{\ib,\jb;\bar{t}}^{\bar{k}}(\bar{z})\ .
\ee
The expansion coefficients
$d_{(i,\ib) (j,\jb)}^{(k,\bar{k});t, \bar{t}}\ $
determine, up to the  normalisation of the chiral blocks,
the  coefficients of the short distance operator product expansion (OPE),
 i.e., these are the {\it relative} OPE coefficients of
 the non-diagonal model if $d$ are chosen trivial for the  diagonal
 model of  same central charge.
   These numbers are constrained by the requirement of locality 
of the physical correlators, which makes use 
of the braiding matrices $B(\pm)$. 
The resulting set of coupled quadratic equations
has been fully solved only  in the $sl(2)$ cases 
(see \cite{DF,PZ1} and further references therein).
\quad

Starting with the ADE  classification of the $sl(2)$ modular
invariants \cite{CIZ}, it has been gradually realised that behind
these 
%spectral and OPE 
data there are hidden {\bf graphs $G$}.  A
systematic study of the graphs generalising the $sl(2)$ ADE
Dynkin diagrams was initiated in \cite{Kos,DFZ1}.  
It was empirically observed that these graphs are encoding a non
trivial information on some spectral and OPE data, more precisely
on those pertaining to the {\it spin-zero} fields of the theory.
These graphs $G_i$, which share the same set of vertices $\calV$,
are labelled by an index $i\in {\cal I}$ --but it is sufficient
to restrict $i$ to a generating subset (generating in the sense
of fusion)-- and their adjacency matrices are commuting and
simultaneously diagonalizable in an orthonormal basis $\{\psi\}$:

\noindent $\bullet $
the  diagonal part of the physical spectrum, 
labelled by the so-called  {\it ``exponents''}\,:
\be 
{\cal  E}=\{j\in  \calI| j=\bar j, Z_{jj}\ne 0\}\ ,
\ee
counted with the multiplicity $Z_{jj}$, is in 
one-to-one correspondence with the spectrum of eigenvalues of $G_i$,
of the form $S_{ij}/S_{1j}$, $j\in \calE$.

\noindent $\bullet$ in the $sl(2)$ (ADE) cases
the structure constants $M_{ij}{}^k$ of the so-called 
Pasquier algebra \cite{Pasq87}
can be identified with the (relative) OPE coefficients
of the scalar fields \cite{PZ1}
\be 
d_{(i,i) (j,j)}^{(k,k)}=M_{ij}{}^k:=\sum_{a\in \calV}\, {\psi_a^i\,
\psi_a^j \,\psi_a^{k\, *}\over \psi_a^1}\,.
\label{strc} 
\ee 
The summation in (\ref{strc}) runs over the set $\cal V$ of vertices
of the Dynkin diagram.

The purpose of this note is to show that the study  
of the CFT on a half-plane or a cylinder %, 
(i.e. on a manifold with boundary), \cite{Ca89},
provides an alternative (chiral) approach in
which  these graphs and the related algebraic structures become
manifest.

%%%%%%%%%%%%%%%%%%%%%%%%%%%%%%%%%%%%%%%%%%%%%%%%%%%%%%%%%%%%%%%%%%%%%%%%%%%%%%%

\section{Graphs and conformal boundary conditions} 
We may summarize the results of \cite{BPPZ} as follows: boundary
conditions that respect the chiral algebra $\goth A$ are
described by a set of commuting, non-negative integer valued
matrices $\{n_i= n_{i a}{}^b\}\,,i\in \calI\,,$ $a,b\in \calV\,,$
$|\cal E|=|\cal V|$, s.t. $n_1=I$, $n_i^T=n_{i^*}$, realising a
representation of the Verlinde fusion algebra
\be 
n_i\;n_j=\sum_{k\in \calI} N_{ij}{}^k\,n_k \,. 
\label{nimrep} 
\ee 
 The matrices $n_i$ thus   admit a spectral decomposition
\be 
n_{i a}{}^b =\sum_{j\in \cal  E}\, {S_{ij}\over S_{1j}}\
\psi_a^j\ \psi_b^{j*}\,,
\label{specdec}   
\ee 
where $\psi_a^j$ are unitary matrices of dimension $|\cal E|=|\cal V|$
and the sum runs over the 
set $\{(j,\alpha)\,, j\in \calI\,,
\alpha=1,2,\dots, Z_{jj}\}$, 
for simplicity of notation identified with
 the set  of exponents $\calE$.
This comes about as follows: On the upper half-plane 
 param\-etrised
by a coordinate $z\in {\Bbb H}^+$,  (resp. on a finite-width
strip $w={L\over \pi}\, \log z\,$), with boundary conditions $b$
and $a$ imposed on the negative and positive real axes, (resp. $\!\!$ on
the two sides of the strip), only one copy of the chiral algebra
acts: only real analytic coordinate transformations,
$\epsilon(z)=\bar{\epsilon}(\bar{z})$ for real $z=\bar{z}$, are
allowed. Thus the Hilbert space of the theory ${\cal H}_{ba}$
splits into a linear sum of representations
\be 
{\cal H}_{ba}=\oplus_{i\in \calI} n_{ib}{}^a \,\calV_i\ .
\label{Hab}
\ee 
On a finite segment of the strip of length $T$ with periodic boundary 
conditions $w\sim w+T$, i.e. on a cylinder, 
the partition functions $Z_{b|a}$ is thus a linear form
in the characters
\be 
Z_{b|a}(\tau)=\sum_{i\in \calI} n_{ib}{}^a\, \chi_i(\tau)\,, \quad
\tau= {i T\over 2 L}\,.
\label{pfn} 
\ee 
The multiplicities $n_{ia}{}^b$ are constrained  by
the Cardy consistency condition  \cite{Ca89}  
which expresses that $Z_{b|a}$ can also be evaluated in a dual way:
by  mapping the cylinder to an annulus region in the full plane, 
one computes $Z_{b|a}$ as the matrix element of the
evolution operator between boundary states 
$|a\rangle$ and $\langle b|$. The latter are decomposed on a standard basis 
of complete and orthonormal ``Ishibashi states'' $|j\rangle\!\rangle$ 
labelled by an exponent $j\in \calE$, 
according to $|a\rangle =\sum_{j\in \cal E}\,  
{\psi_a{}^j \over \sqrt{S_{1j}}}
\ |j\rrangle $. This gives
\bea
Z_{b|a}(\tau)
&=&
\langle b|e^{{-\pi i\over \tau} (L_0+\overline{L}_0-{c\over 12})}
|a\rangle
\nonumber\\ 
&=&
 \sum_{j\in \calE}{\psi_a{}^j \psi_b{}^{j*} \over {S_{1j}}}
\chi_j\Big({-1\over\tau}\Big)\ . \label{cardy}  
\eea 
Comparing the resulting two expressions for %\\
  $Z_{b|a}$ 
yields the
spectral decomposition (\ref{specdec}). See \cite{BPPZ} for a discussion of
the assumptions and a more detailed derivation. 

Each set of matrices  $\{n_i\}$ solving (\ref{nimrep}), (\ref{specdec})  
may be regarded as the adjacency matrices of a collection of graphs
$G_i$, thus explaining the occurence of graphs with the special 
spectral properties noticed above.
Solving in general  the system  (\ref{nimrep}) 
provides   a classification of boundary conditions,   labelled by
the vertices of the graphs $G_i$ 
and specified by the spectrum (\ref{Hab}).  
The  case $n_i=N_i $ provides the regular
 representation with spectrum $\cal  E=\cal  I$ corresponding to the
 diagonal case;
then (\ref{nimrep}) reduces to (\ref{Verl}) with $\psi=S$
and the boundary states are labelled by the same indices $i\in \calI$ as 
the representations and primary fields \cite{Ca89}.

$\bullet$
 In $\slh(2)$ RCFT  
(the WZNW models and the Virasoro minimal models)
this reduces to the classification of the symmetric, irreducible, non-negative
integer valued matrices of spectrum $|\zg^j|$ 
$ =|{S_{2 j}\over S_{1 j}}|< 2$.
This is well known to lead to an ADE classification, after
the  spurious ``tadpole'' graphs  $A_{2n}/{\Bbb Z}_2$ have been discarded, 
on the basis that their spectrum does  not match any known modular
invariant.

$\bullet$
A new situation arises in the $sl(3)$ case, where 
there is no known a priori classification of (oriented) graphs 
with the required spectral properties. Solutions may be found
\cite{DFZ1}
and comparison with the complete list of modular invariants \cite{Ga}
enables one to discard some spurious solutions and leads to 
a list of graphs, see \cite{BPPZ}. 
One also finds that in a few cases, there are isospectral graphs, i.e.
 more than one solution with a given spectrum, indicating that 
there are several choices of boundary conditions
 (and in fact of OPE  structure constants, see \cite{BPPZ}), 
for a given spectrum of  spin-zero fields in a RCFT.
%
%
%%%%%%%%%%%%%%%%%%%%%%%%%%%%%%%%%%%%%%%%%%%%%%%%%%%%%%%%%%%%%%%%%%%%%%%%

\section{ Ocneanu quantum graph symmetry} 

According to Cardy the boundary conditions $(a,b)$ are
created by insertions of 
 fields on the boundary ({\it boundary fields}),
  $^a\Psi_{j;\beta}^b(x)$,
$\beta =1,2,\dots n_{jb}{}^a\,,\ $ $x\in {\Bbb R}$.
%, $z\in {\Bbb H}^+$.
Exploiting some ideas of Ocneanu \cite{Ocn,Ocn00}
one can interpret these fields, extended to the full plane, as 
generalised CVO.

  Given a solution of the system of equations
   (\ref{nimrep}), consider 
for each $j\in \cal I$ an auxiliary  Hilbert space 
$V^j\cong {\Bbb C}^{m_j}\,$ of dimension $m_j= \sum_{a,b}\,
n_{ja}{}^b$ with basis states $|e^{j,\beta}_{ba}\rangle$, 
%					       |j
%					    b__|__a
$\beta=1,2,\dots,n_{j a}{}^b$, depicted as  
\hbox{\raise -7mm\hbox{\epsfxsize=3cm\epsfbox{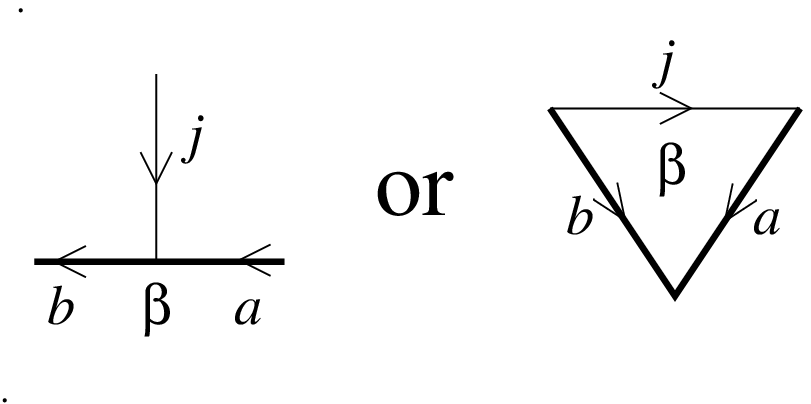}}}.
A scalar product in $\oplus_{j\in \calI}\, V^j$ is defined as
\bea 
 \langle e^{j,\beta}_{ ab} |
e^{j',\beta'}_{a'b'}\rangle 
&=& \delta_{bb'} \delta_{aa'}\,
\delta_{jj'}\,\delta_{\beta' \beta}\,\sqrt{P_a\,P_b\over  d_j}
\,, \nonumber \\
d_j&:=&{S_{j1}\over S_{11}}\,,\ P_a:={\psi_a^1\over \psi_1^1}\,. \label{ebasis}
\eea 
Restricting to a subspace $V^i \otimes_h\, V^j\,$ of $V^i\otimes V^j\,,$
with coinciding intermediate labels, one defines a decomposition
$V^i\otimes_h\, V^j\cong \oplus_{k}\, N_{ij}{}^k\,V^k$,
  or, explicitly,
\bea 
\vert e^{i, \eta}_{ab}\ket \otimes_h \vert e^{j,\zeta}_{b c}\ket &=&
\, \sum_{k\in \calI }\ \sum_{\gamma=1}^{n_{kc}{}^a}\,
\sum_{t=1}^{N_{ij}{}^k}\  
\Fo_{b k}\left[ {i\atop a}\ {j\atop c} \right]_{\eta\, \zeta}^{\gamma\, t}\,
\nonumber \\
% & &\!\!\!\!\! \!\!\!\!
\times \sqrt{P_b}\, 
& &\!\!\!\!\! \Big({d_k \over d_i d_j}\Big)^{{1\over 4}}\
|e^{k\,, \gamma}_{ac}(ij;t)\ket\,.
\label{prodh} 
\eea 
\omit{Pictorially \hbox{\raise -4mm\hbox{\epsfxsize=4cm\epsfbox{Fone.eps}}} 
%	|    |	               \/
%	|    |		        |			Fig 2
%      ________  = sum ^1F  _________
}
%%%%
The counting of states in both sides is consistent
 due to (\ref{nimrep}).
 The  Clebsch-Gordan coefficients $\Fo\in {\Bbb C}$
({\bf ``$3j$- symbols''})  satisfy
 unitarity conditions (completeness and orthogonality)
and a pentagon identity,  reflecting the  associativity of the
product (\ref{prodh}), written schematically as
\be 
 F\  \Fo \ \Fo \, =   \Fo\ \Fo \,,\qquad 
\label{pentagon} 
\ee 
where $F$ are the $6j$-symbols. In the diagonal case $F$ and $\Fo$ 
are identified, and (\ref{pentagon})  becomes the conventional pentagon 
identity satisfied by the fusing matrices of the corresponding RCFT.

The space  $\calA =\oplus_{j\in \calI}\, \hbox{End}\,(V^j)$
is a matrix algebra $\oplus_{j\in \calI}\, M_{m_j}$.
A  matrix unit basis
$\{e_{j;\eta\,,\eta'}^{(ab)\,,(a'b')}\\
=| e^{j,\eta}_{ab}\ket \bra\,e^{j,\eta'}_{a'b'} |
\,, j\in\cal I\}$ in $\calA$ is identified
with a product of states in $V^j\otimes V^{j^*}$,
and depicted as $4$-point blocks or double triangles  \\
%\medskip
% <<<  Fig
\qquad\hbox{\raise -4mm\hbox{\epsfxsize=5cm\epsfbox{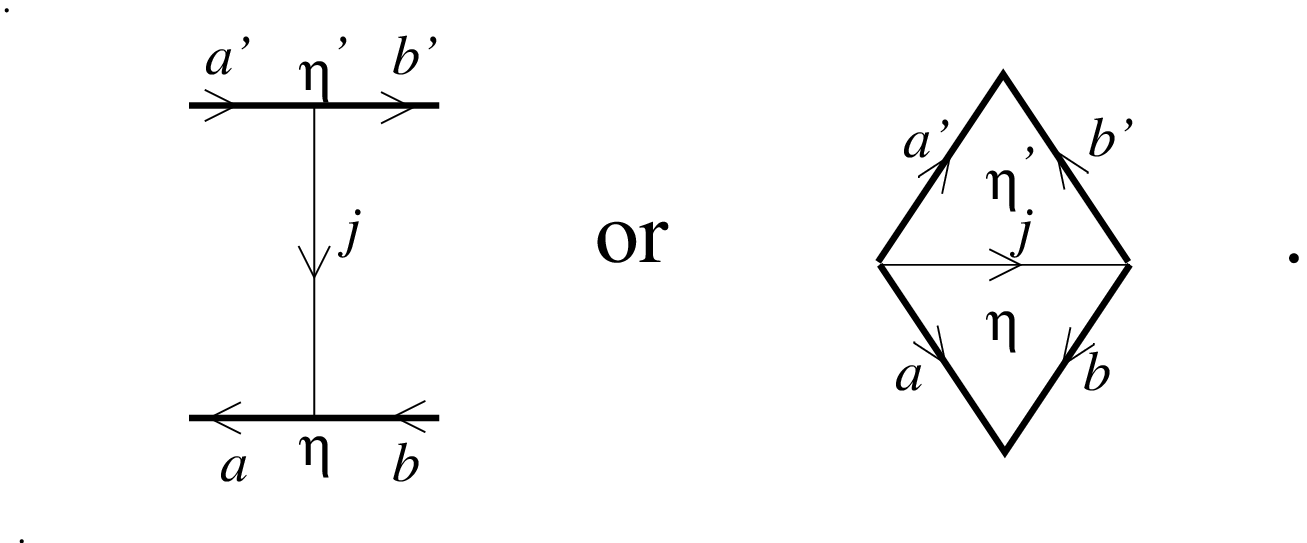}}}\\ 
\noindent
 Along with the matrix (``vertical'') product, 
 a second ``horizontal'' product
 (or, alternatively, a coassociative coproduct), is defined   on $\cal A$
 via the $3j$-symbols $\Fo\,$
 \\ \noindent
\hbox{\raise -4mm\hbox{\epsfxsize=7cm\epsfbox{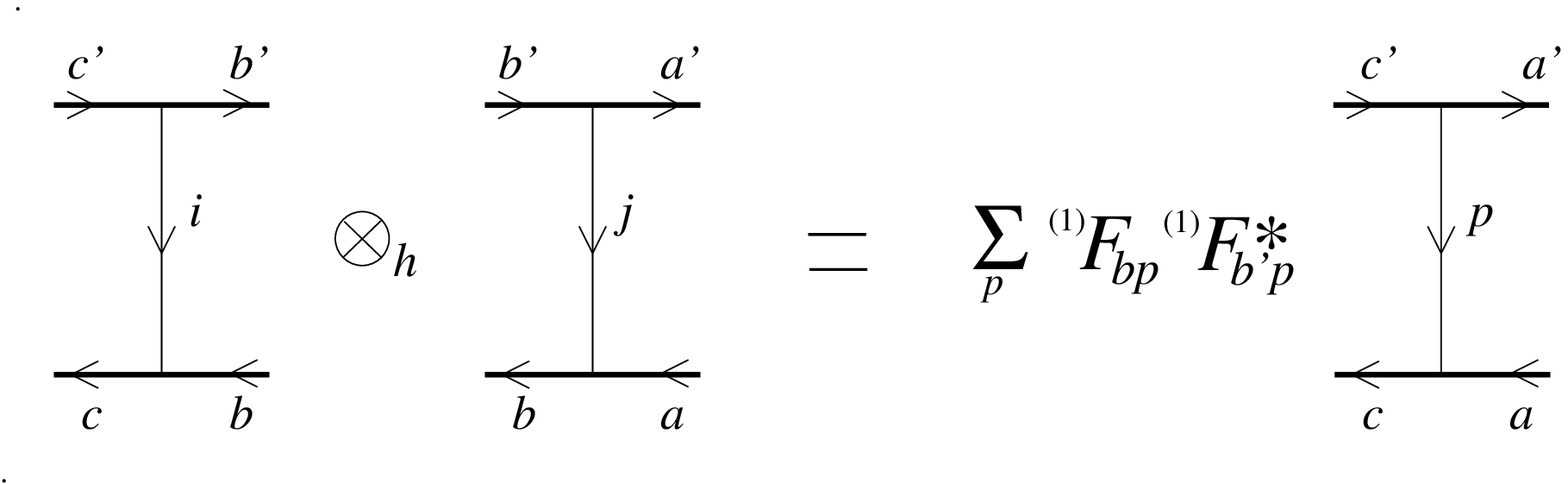}}}\\ 
\noindent
The algebra generated by these objects
 is the Ocneanu {\bf double triangle algebra} (DTA) \cite{Ocn},
further studied in the subfactor theory, see \cite{BEK,BE} and 
references therein.
Equipped with a counit and an antipode satisfying a weakened version
of the Hopf algebra axioms ($\triangle( 1_v)\ne 1_v\otimes 1_v$, etc.),
  it was considered in  \cite{BoSz} as an example of   a
{\it weak $C^*$- Hopf algebra} (WHA). This  Ocneanu ``graph
quantum algebra'',   
associated to any solution $\{n_i\,,i\in \calI \}$ of $(\ref{nimrep})$,  
together with its dual algebra $\hat{\calA}$ structure (see below), 
appears as the quantum symmetry of the 
 CFT model,  either diagonal or non-diagonal. 
%%%%%%%%%%%%%%%%%%%%%%%%%%%%%%%%%%%%%%%%%%%%%%%%%%%%%%%%%%%%%%%%%%%%%%%

\section{Generalised CVO (GCVO)}

We define operators  
$\oplus_{j\in \calI}\, \calV_j\!\otimes\! V_j \!\to\!
\oplus_{k\in \calI}\, \calV_k\!\otimes\! V_k$\\
\noindent associated with the basis states in the tensor product
(\ref{prodh}), 
\bea 
^a\Psi_{i, \beta;I}^c(z) 
&=&\sum_{j,k, t}\ \phi_{ij;t,I}^k(z)  \otimes\,
 \sum_{ b,\alpha,\gamma}\,
\Fo_{ck}\left[{i\atop a} {j\atop b}\right]^{ \alpha\, t}_{
\beta\, \gamma}\, \nonumber \\
& &  \times \sqrt{d_j\over P_c\,P_b} \
| e^{k,\alpha}_{ab}\ket \bra\,e^{j,\gamma}_{cb} | \,,
\label{gcvo}  
\eea 
covariant under the action of $\calA$.
Their correlators, obtained by projecting on a state
$|0\rangle \otimes |e^1_{aa}\rangle  $ in the vacuum 
space $\calV_1\otimes V^1$,   as well as 
 their fusing  and braiding properties, are inherited 
 from those of the conventional CVO.
 In contrast with previous approaches  introducing quantum 
 symmetry covariant operators, e.g.,  \cite{MR, MSch}, note that in
the diagonal case the $6j$-symbols $F$  of the
related quantum group (instead of its $3j$-symbols)
 appear in the r.h.s. of (\ref{gcvo}).
 For real $z$ the operators  $\Psi$ of (\ref{gcvo})
represent (after an appropriate choice of normalisation)
the boundary fields, and their short  distance 
$x_{12}\approx 0$  expansion is recovered, with the $3j$-symbols 
 $\Fo$ serving as {\it boundary fields OPE coefficients}.  
The recoupling equation (\ref{pentagon}) expresses then the
associativity of the boundary fields OPE  \cite{BPPZ} and  is
equivalent to the  Lewellen boundary fields sewing identity
\cite{CL}. 
 
 In general the chiral operators (\ref{gcvo}) have nontrivial
braiding with a new braiding matrix $\hat{B}(\epsilon)$, with
$4+2$ indices of two types,
\bea  
& &^a\Psi_{j,\alpha}^b(z_1)\, ^b\Psi_{k,\gamma}^c(z_2)= 
\label{newbraid} \\
& &\sum_{d,
\alpha',\gamma'}
\hat{B}_{bd}\left[{j \atop a} {k\atop c} \right]_{
\alpha\, \gamma}^{\alpha'\, \gamma'}\!\!\!\!(\epsilon)\
^a\Psi_{k,\alpha'}^d(z_2)\, ^d\Psi_{j,\gamma'}^c(z_1)\,,
\nonumber 
\eea  
$z_{12}\notin {\Bbb R}_- \,, \
\epsilon=\hbox{sign}\,(\hbox{Im}\,z_{12}$). 
The matrices $\hat{B}$ satisfy various relations:
 
$\bullet$
 Inversion (unitarity) relation
\be  
\hat{B}^{12}(\epsilon) \, \hat{B}^{21}(-\epsilon)=1\,.
\ee  

 $\bullet$
 The braid group (``Yang-Baxter'') relation %equation 
\be  
\hat{B}^{12}\, \hat{B}^{23}\,  \hat{B}^{12}\,
=  \hat{B}^{23}\, \hat{B}^{12}\,  \hat{B}^{23} \,.
\label{braidgp}
\ee  

 $\bullet$  The braiding--fusing (pentagon) identity
\be  
 \hat{B} \ \Fo =  \Fo\  \hat{B}\  \hat{B}\,,
\ee  
which turns into a recursion relation for $\hat B$,
 given  the fundamental  $3j$-symbols,

  $\bullet$
Intertwining relation
 \be  
\Fo\, \Fo\ B= \hat{B}\ \Fo\, \Fo\,, \label{intwg}
\ee  
which implies a bilinear representation of $\hat{B}$ 
in terms of the $3j$-symbols  
\bea 
& &  \hat{B}_{bd}\left[{j\atop a} {k\atop c} \right]_{
\alpha \gamma}^{\alpha' \gamma'}\!\!\!\!\!\!(\epsilon)
 =
  \label{bilin}\\
 & & \sum_{i\,, \delta\,, t }\,
{}\Fo_{b i}\left[{j\atop a} {k\atop c}\right]_{\alpha
\gamma}^{\delta\,  t }
 e^{-i \pi\epsilon\, \triangle_{jk}^i}\
{}\Fo_{d i}^*\left[{k\atop a} {j\atop c} \right]_{\alpha'
\gamma'}^{\delta\,  t }\,,
\nonumber 
\eea  
where $\triangle_{jk}^i$  is the combination of
  conformal weights $\triangle_j+\triangle_k-\triangle_i $.
%In the diagonal case  $\hat{B}(\epsilon)$ is identified
%with $B(\epsilon)$.

%%%%%%%%%%%%%%%%%%%%%%%%%%%%%%%%%%%%%%%%%%%%%%%%%%%%%%%%%%%%%%%%%%%%%%

\section{Connection to lattice models}
Some of these identities have been already encountered in
critical lattice models, the  ADE Pas\-quier models and their
$\widehat{sl}(n)_{h-n}$ generalisations, in which the degrees of
freedom may be regarded as vertices of the previous graphs
\cite{Pa1,DFZ1}.

In the limit $u\to -i\epsilon\infty$, ($\epsilon=\pm1$), of the
spectral parameter $u$,  the face  Boltzmann weights  $W(u)$
satisfy the same equation (\ref{braidgp}) as the braiding matrix
$\hat{B}$ of (\ref{newbraid}).  Indeed, denoting the
representations of $sl(n)$ by their Young tableau, and with
$q=e^{2\pi i\over h}\,$,
\bea  
& & \hat{B}_{bd}  
\left[{\youngone\atop a}{\youngone\atop c} \right]_{
\za\, \zg}^{\za'\, \zg'}\!\!\!\!(\epsilon)\, 
= \nonumber \\
& &  2i \, q^{-\epsilon {1\over 2 n}}\, \lim_{u\to -i\epsilon\infty}\,
e^{-i\pi\epsilon u}\, W_{bd}\left({a\atop c} \right)_{
\alpha\, \gamma}^{\alpha'\, \gamma'} (u)\,,
 \nonumber 
\eea  
\be  
W_{bd}(u)=\sin({\pi\over h}-u) \delta_{bd} +\sin(u)\, [2]_q\, U_{bd}
\,,\quad U^2=U\,.
 \ee 
Using (\ref{bilin}), 
the Hecke algebra generators $U$ are expressed via the $3j$-symbols,
recovering an Ansatz of Ocneanu \cite{Ocn00} 
\be  
U_{bd}=\sum_{\zb}\, 
\Fo_{b\, \youngtwo}\left[{\youngone\atop a} {\youngone\atop
c}\right]_{\za\, \zg}^{\zb\ 1}\, 
\Fo_{d\, \youngtwo}^*\left[{\youngone\atop a}
{\youngone\atop c} 
\right]_{\za'\, \zg'}^{\zb\ 1 }\,.
\ee  

Moreover, equation (\ref{intwg}) shows that we can identify  
the $3j$- symbols  $\Fo$  with the 
{\it ``intertwining
 cells}''   studied in these models, see e.g. \cite{DFZ1}; then
 lattice results provide solutions for particular $\Fo$, namely, those
for which one of the labels in $\cal I$ is equal to $\youngone$.

For $sl(2)$: $\youngtwo=1\,,\ $
 $ [2]_q\ U_{bd}  = \delta_{ac}\,{\sqrt{P_b\,P_d}\over P_a }$, while 
for $sl(3)$: $\youngtwo=\youngone^*\,,$ and  these cells $\Fo_{b\,\youngone
{}^*}$
exist for all graphs listed in \cite{BPPZ} but one, according to   
the recent work of  Ocneanu, \cite{Ocn00}.
% completing previous results of {\bf Jimbo et al, Fendley, DiFZ, Sochen}).

%%%%%%%%%%%%%%%%%%%%%%%%%%%%%%%%%%%%%%%%%%%%%%%%%%%%%%%%%%%%%%%%%%%%%%%%%

\section{Bulk fields}

For $(j,\bar{j})$ in the physical spectrum (\ref{Spec}), 
i.e., $Z_{j\bar{j}}\not =0$, and
$z\in {\Bbb H}^+\,,$ define  (upper) half-plane bulk fields 
by two-point compositions of 
%generalised CVO (\ref{gcvo})
GCVO (\ref{gcvo})
\be 
\Phi_{(j,\bar{j})}^H(z, \bar z)
=\sum_{a,b\,,\atop{\beta'\,,\beta}}\
C^{a,b,a;\beta,\beta'}_{(j,\bar{j})}\
{}^a\Psi_{j, \beta}^b(z)\ ^b\Psi_{\bar{j}^*,
\beta'}^a(\bar{z}) \,.
\label{bulkf}  
\ee 
{}From this operator representation, which
can be also rewritten
%, using (\ref{gcvo}),
 in terms of the conventional CVO, one  
recovers the bulk field
small distance $z-\bar{z}=2i\,y\approx 0$ expansion,
in the leading order 
\bea
& & 
%\bra e^1_{aa}|
\Phi_{(j,\bar{j})}^H(z, \bar z) =
\label{bulkbdy} \\ & & 
%|e^1_{aa}\ket
%=
\!\!\!\sum_{p, a, \za, t } R_{(a;\za)}^{(j,\bar{j};t)}(p)\,
%\label{bulkbdy} \\ & & 
\bra  p| \phi_{j\bar{j}^*; t}^p(2 i y)|\bar{j}^*\ket\, 
%\bra e^1_{aa}|
{}^{a}\Psi_{p, \za}^a(x)
%|e^1_{aa}\ket
 +   \dots \,,
\nonumber 
\eea
with $R=\sum \, \Fo\ C $ -- the bulk-boundary  reflection
coefficients  of Cardy--Lewellen \cite{CL}.  Given the chiral
representation (\ref{bulkf}) all the bulk-boun\-da\-ry  sewing
relations of \cite{CL} are derived from the duality
identitities of the conventional CVO.

$\bullet$
Diagonal case: the two basic bulk-boundary Lewellen equations 
are not independent and are 
equivalent to the torus duality identity of
Moore-Seiberg with  $R(p) \sim S(p)$ \cite{BPPZ}.
%%%%%%%%%%%%%%%%%%%%%%%%%%%%%%%%%%%%%%%%%%%%%%%%%%%%%%%%%%%%%%%%%

\section{The dual DTA and the physical spectrum}

The dual (in the algebraic sense)
of Ocneanu DTA leads to the consideration of new  graphs
$\tilde{G}$, \cite{Ocn,Ocn00},
(see also  \cite{BE, BEK}),
with a set of vertices
$ x \in \tilde{\calV}=\{1,2,\dots, \sum_{i,j}\, (Z_{ij})^2\}$. %, 
These graphs are described by  
a set of nonnegative integer
valued matrices    $\big(\tilde{V}_{ij}\big)_x^{\ y}$, 
forming a representation of the  product of two  Verlinde
algebras, thus generalising (\ref{nimrep})
\be
 \tilde{V}_{i_1j_1} \tilde{V}_{i_2j_2}=
\sum_{i_3,j_3} N_{i_1i_2}{}^{i_3} N_{j_1j_2}{}^{j_3}\
\tilde{V}_{i_3j_3} 
\ee
with the additional condition that
$\big(\tilde{V}_{ij^*}\big)_1^{\ 1}= Z_{ij}$, the matrix of 
(\ref{partfn}).  The eigenvalues of
$\tilde{V}_{ij}$ are described by this matrix $Z$,
i.e., are labelled by the set $\tilde{\calE} =
\{ (\ell,\bar\ell)\,, \ell,\bar\ell\in \calI\,,$ taken
with multiplicity $(Z_{\ell\bar\ell})^2\} $,
$|\tilde{\calE}| = |\tilde{\calV}|$  , 
and are of the form 
 $S_{i\ell}S_{j\bar\ell}/S_{1\ell}S_{1\bar\ell}$. 
For example in the $E_6$ case, this graph has 12 vertices
that we label by a pair $x=(a,b)$, $a=1,2,\cdots,6$, $b=1,2$
of two vertices of the ordinary $E_6$ diagram; the new graph is
generated by ${\tilde V}_{2,1}$ and    ${\tilde V}_{1,2}$
and the corresponding edges are depicted in red and blue 
respectively on the following figure 
due to Ocneanu \\
\qquad{\epsfxsize=5cm\epsfbox{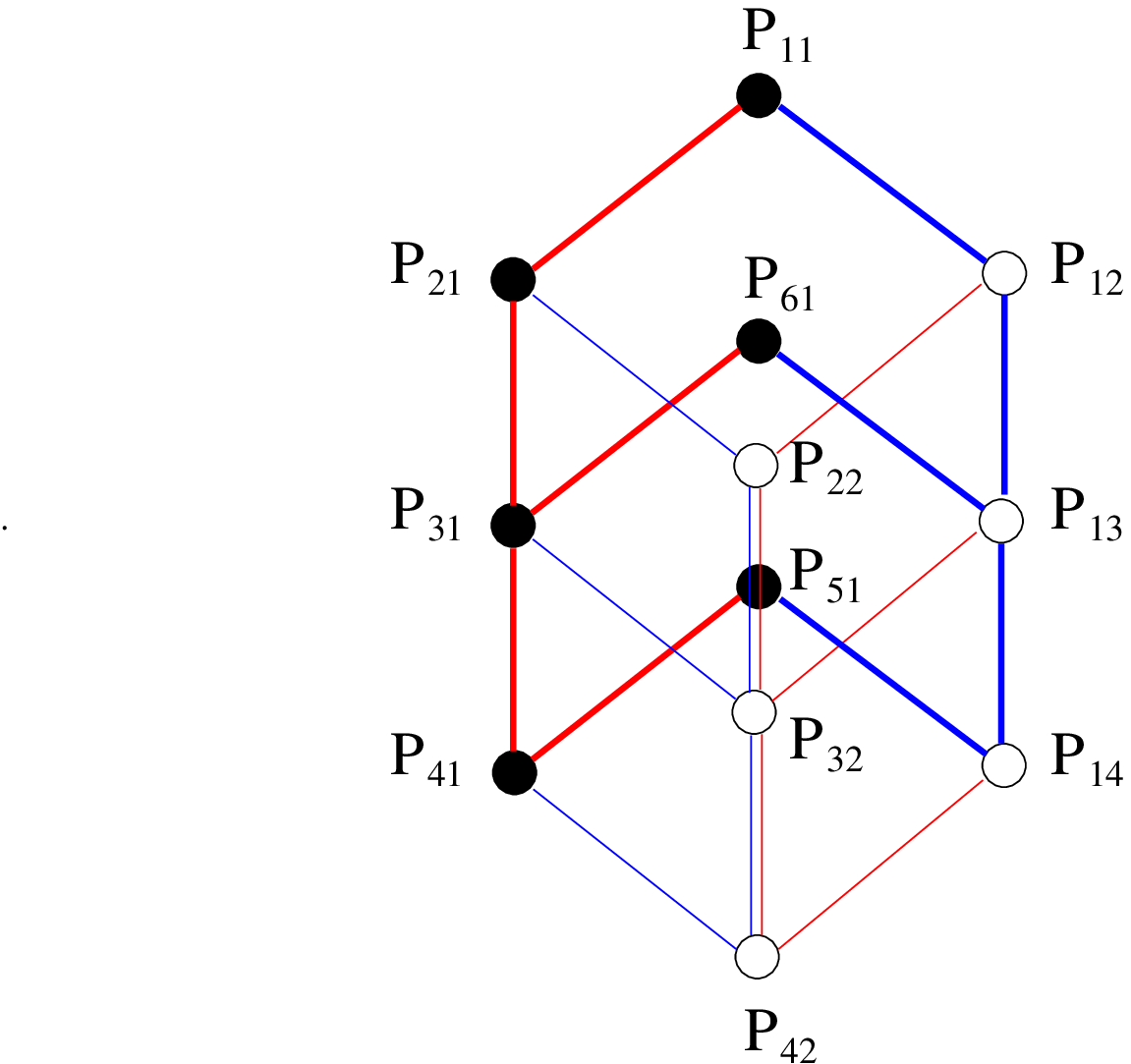}}

\noindent
On that figure $ (P_{ab})_{ij}:=\sum_{c=1,5,6} n_{ia}{}^c
n_{jb}{}^c$, expressed in terms of $n_i$ of (\ref{nimrep}), give
the  matrices $({\tilde V}_{ij})_1{}^x$, generalising a well
known formula for $Z=P_{11}$
\cite{DFZ2,PZ2}.

This Ocneanu graph $\tilde{G}$ gives rise to a (noncommutative in
general) fusion algebra with structure constants
$\tilde{N}_{xy}{}^z$; the nodes of the graph can be equivalently
associated with the matrices ${\tilde N}_x$. This algebra also
admits a representation by matrices $(\tilde{n}_x)_a^b=
\tilde{n}_{xa}{}^{b}$ of size $|\calV|$ with nonnegative integer
entries
\be  
\sum_{b\in \calV}\, \tilde{n}_{xa}{}^b \, \tilde{n}_{yb}{}^c=
\sum_{z \in \tilde{\calV}}\,
\tilde{N}_{xy}{}^z\,
\tilde{n}_{za}{}^c \,.
\ee  
These integers describe the dimensions $\tilde{m}_x=\sum_{a,b}\,
\tilde{n}_{xa}{}^b$ of the dual spaces $\hat{V}_x$, spanned by
states (dual triangles) $|E_{a x;\za}^b\ket\,,$ $\za =1,2,\dots,
\tilde{n}_{ax}^{b}$.  The ``dual double triangles'' span a basis
of  $\hat{\calA}$, with vertical and horizontal products
reversed. The two basis sets are related by a ``fusing'' matrix,
$\Fos$, satisfying the  pentagon identity
\be  
\Fos \ \Fos = \Fos \ \Fo\ \Fo^*\,.
\ee  
More pentagon relations involve the dual  $3j$- and $6j$-symbols;
the full set of  such relations is  named the  ``Big Pentagon
identity'' in \cite{BoSz}.  In the diagonal cases, all these $F$
matrices coincide.  The problem is to extend this general
construction beyond the diagonal case.

{\bf Example:} $sl(2)$ $D_r$ series, $r={h\over 2}+1$ odd.   
Here
$\tilde{\calV}=\{x\equiv  k =1,2,\dots, h-1\}\,,$
$\tilde{\calE}=\{(l, \zeta(l))\,  |\,\\ l=1,2,\dots, h-1\}\,,$
%fusion rules preserving automorphism  
where $\zeta(l)=h-l$ for $l$ even, $\zeta(l)=l$ for $l$ odd.
Then
$({\tV_{ij}})_1{}^k=N_{i \zeta(j)}{}^k$
is diagonalised by  $S$ and $\tilde{G}= A_{h-1}\,$,
$\tilde{N}_k=N_k\,,$
  $\tilde{n}_k= n_k$. Hence
$\cal A$ is  self-dual,  with both the $3j$- and $6j$-symbols
being self-dual. The $3j$-symbols $\Fo$ were found in \cite{Ru}
and it remains to determine the $ \Fos$ 
to  have a non-diagonal example of WHA.
\medskip

While $\calA$ gives meaning to the generalised CVO, the CFT
interpretation of the  dual $\hat{\calA}$ is less clear.  The
representations of the  $\tilde{N}$-algebra carry the labels
$(j,\jb)$ of  the physical fields with $Z_{(j,\jb)}\ne 0$.  In
the cases $Z_{ij}=0,1$, when this algebra is commutative with
only 1-dimensional representations,  one can consider its dual
algebra, a generalised Pasquier algebra, with structure constants
\be
\tilde{M}_{(i,\bar{i}) (j,\bar{j})}^{(k,\bar{k})} =
\sum_{x\in \tilde{\calV}}\,
{\Psi_x^{(i,\bar{i})}\over \Psi_x^{(1,1)}}\,
\Psi_x^{(j,\jb)}\,
\Psi_x^{(k,\bar{k})*}\,,
\ee
where  $\Psi_x^{(i,\bar{i})}$ is a unitary matrix diagonalising
$\tilde{V}_{ij}$ and $\tilde{N}_x$.
In all
$sl(2)$ cases of this type, i.e., $A$, $D_{odd}$,
$E_6\,,E_7\,,E_8\,,$ 
one finds \cite{PZ4}  a relation generalising (1.4),
\be
\quad
\tilde{M}_{(i,\bar{i}) (j,\bar{j})}^{(k,\bar{k})}=
\Big (d_{(i,\bar{i}) (j,\bar{j})}^{(k,\bar{k})} \Big)^2\,,
\label{gP}
\ee
involving
the relative OPE  coefficients  of  fields of
non-zero spin. For $E_6, E_8\,,  $ one has
$
\tilde{M}_{(i,\bar{i}) (j,\bar{j})}^{(k,\bar{k})}=
M_{ij}^k\ M_{\bar{i} \bar{j}}^{\bar{k}}\,,
$
\noindent
in agreement with the known factorisation of the OPE coefficients
in these cases.  The problem remains to recover  (\ref{gP})
directly in the field theory framework  as it has been achieved
in the boundary CFT \cite{PSS, BPPZ} for its scalar counterpart
(\ref{strc}).

\bigskip

%%%%%%%%%%%%%%%%%

\noindent{\bf Acknowledgements:~} \\ \noindent
Our warmest thanks to Adrian Ocneanu for his generous
explanations of his recent work prior to publication.
We also thank Paul Pearce and Peter Vecsernyes for useful
discussions.
V.B.~P. acknowledges partial support of the Bulgarian National
Research Foundation (Contract $\Phi$-643). J.-B.~Z.  acknowledges
partial support of the EU Training and Mobility of Researchers
Program (Contract FMRX-CT96-0012).

%%%%%%%%%%%%%%%%%%%%%%%%%%%%%%%%%%%
%
%\begin{thebibliography}{99}

%%%%%%%%%%%%%

\end{document}